\documentclass[11pt]{iopart}
\usepackage{iopams}  
\usepackage{epsf}

\def\Journal#1#2#3#4{{#1} {\bf #2}, #3 (#4)}

\def\CQG{\em Class. Quantum Grav.}
\def\PRD{\em Phys. Rev. D }
\def\GRG{\em Gen. Rel. Grav.}

\def\PRL{\em Phys. Rev. Lett.}

\def\PREP{\em Phys. Rep.}

\def\PLB{\em Phys. Lett. B }
\def\NPB{\em Nuc. Phys. B }
\def\NC{\em Nuovo Cimento }
\def\APNY{\em Ann. Phys. (N.Y.)}

\newcommand{\bm}[1]{\mbox{\boldmath $#1$}}
\def\d{{\rm d}}

\def\espaitemps{({\cal V},g)}
\def\varietat{{\cal V}}

\def\r{{\rm I\!R}}
\def\di{{\rm div}}
\def\sg{{\rm sign}}

\def\doo{d \Omega^2_{D-2}}

\def\be{\begin{equation}}
\def\ee{\end{equation}}
\def\bea{\begin{eqnarray}}
\def\eea{\end{eqnarray}}
\def\bean{\begin{eqnarray*}}
\def\eean{\end{eqnarray*}}

\begin{document}
\setcounter{footnote}{2}
\title{Trapped surfaces, horizons and exact solutions in
higher dimensions}
\author{Jos\'e M. M. Senovilla}
\address{Departamento de F\'{\i}sica Te\'orica,
Universidad del Pa\'{\i}s Vasco,
Apartado 644, 48080 Bilbao, Spain.}


\begin{abstract}
A very simple criterion to ascertain if $(D-2)$-surfaces are {\it trapped}
in arbitrary $D$-dimensional Lorentzian manifolds is given.
The result is purely geometric,
independent of the particular gravitational theory, of any field equations
or of any other conditions. Many physical applications arise,
a few shown here: a definition of general horizon, which
reduces to the standard one in black holes/rings and other known cases;
the classification of solutions with a $(D-2)$-dimensional abelian group of 
motions and the invariance of the trapping under simple dimensional reductions
of the Kaluza-Klein/string/M-theory type. Finally, a stronger result
involving {\it closed} trapped surfaces is presented. It provides in
particular a simple sufficient condition for their absence.
\end{abstract}

PACS Numbers: 04.50.+h, 04.20.Cv, 04.20.Jb, 02.40.Ky

\vspace{5mm}
In 1965 Penrose \cite{P2} introduced in General Relativity (GR)
the concept of {\it closed trapped surface}, which was crucial
for the development of the
singularity theorems and the study of gravitational collapse, black
holes, cosmological expansion and several types of horizons,
see e.g.\ \cite{HE,S}. Trapped
surfaces (closed or not) are 2-dimensional imbedded spatial surfaces such that
any portion of them has, at least initially, a decreasing area along 
{\it any} future evolution direction. The term ``closed'' is used if
the surfaces are compact without boundary \cite{P2,HE,S}.

This concept carries over
to general Lorentzian manifolds $\espaitemps$ of any dimension $D$ 
\cite{Kr}.
To fix ideas and notation, let $S$ be a $(D-2)$-dimensional
surface with intrinsic coordinates $\{\lambda^A\}$ ($A,B,\dots =2,\dots
,D-1$) imbedded into the spacetime
$\varietat$ by the parametric equations
\begin{equation}
x^{\alpha} =\Phi ^{\alpha}(\lambda^A) \hspace{1cm}
(\alpha,\beta \dots =0,1,\dots ,D-1).\label{imm}
\end{equation}
$S$ is alternatively locally
defined by two independent relations $F_1(x^{\mu})=0$ and $F_2(x^{\mu})=0$.
The tangent vectors $\{\vec{e}_A\}$ of $S$ are
\bean
\vec{e}_A \equiv e^{\mu}_A \left.\frac{\partial}{\partial x^{\mu}}
\right\vert _S \equiv
\frac{\partial \Phi^{\mu}}{\partial \lambda^A}
\left.\frac{\partial}{\partial x^{\mu}}\right\vert_S
\eean
so that the first fundamental form of $S$ in $\varietat$ reads
\begin{equation}
\gamma _{AB}\equiv \left.g_{\mu \nu}\right\vert_S\frac{\partial \Phi^{\mu}}
{\partial \lambda^A}\frac{\partial \Phi^{\nu}}{\partial \lambda^B} \label{fff}
\end{equation}
which gives the scalar products of the $\{\vec{e}_A\}$ in $\espaitemps$.
Assume that $\gamma _{AB}$ is positive definite so that $S$ is
{\it spacelike}. Then, the two linearly independent normal one-forms
$k_{\mu}^{\pm}$ to $S$ can be chosen to be null and future directed
everywhere on $S$, so they satisfy
\begin{equation}
k_{\mu}^{\pm}e^{\mu}_A=0,\hspace{2mm}  k^{+}_{\mu} k^{+ \mu}=0 ,\hspace{2mm}
 k^{-}_{\mu} k^{- \mu}=0 ,\hspace{2mm}  k_{\mu}^+ k^{-\mu}=-1 , \hspace{4mm}
\mbox{on $S$}\label{k}
\end{equation}
where the last equality is a condition of normalization.
Obviously, there still remains the freedom
\begin{equation}
k^+_{\mu} \longrightarrow k'^+_{\mu}=\sigma^2 k^+_{\mu}, \hspace{1cm}
k^-_{\mu} \longrightarrow k'^-_{\mu}=\sigma^{-2} k^-_{\mu} \label{free}
\end{equation}
where $\sigma^2$ is a positive function defined on $S$.

The two null (future) second fundamental forms of $S$ are given by
\begin{equation}
K^{\pm}_{AB} \equiv -k^{\pm}_{\mu}e^{\nu}_A\nabla_{\nu}e^{\mu}_B  \label{sff}
\end{equation}
and their traces are
\begin{equation}
K^{\pm} \equiv \gamma^{AB}K^{\pm}_{AB}, \label{tr}
\end{equation}
where $\gamma^{AB}$ is the contravariant metric on $S$:
$\gamma^{AC}\gamma_{CB}=\delta^A_B$. The scalar defining
the trapping of $S$ is then
\begin{equation}
\kappa \equiv 2\, K^+K^- =H_{\mu}H^{\mu}\label{kappa}
\end{equation}
where $\vec{H}\equiv -K^-\vec{k}^+ - K^+\vec{k}^-$ is the mean curvature 
vector of $S$ \cite{Kr}. Clearly, $\kappa$ and $\vec H$ are invariant under
(\ref{free}). $S$ is said to be {\it trapped}
(respectively {\it marginally trapped, absolutely non-trapped})
if $\kappa$ is positive, (resp.\ zero, negative) everywhere on $S$.
$S$ is called {\it untrapped} otherwise.
See \cite{HE} and section 4 in \cite{S} for details and
examples. Notice that $S$ is trapped (resp.\ absolutely non-trapped)
when $\vec{H}|_{S}$ is timelike (resp.\ spacelike). If $\vec H$
is null at a point, then at least one of the traces $K^{\pm}$
vanishes there, so that a necessary condition for $S$ to be
marginally trapped is that $\vec{H}|_{S}$ be null.

The meaning of the trapping is simple:
the traces (\ref{tr}) are in fact the
expansions of the two families of null geodesics emerging
orthogonally from $S$, which are tangent to
$\vec{k}^{\pm}$ at $S$ \cite{HE,S}. Thus, $S$ is trapped
if both null geodesics families are converging, or diverging, all over $S$.

In this letter, a very simple way to check the trapping of surfaces
is found, and thereby a definition of horizon will also arise naturally.
Without loss of generality, the family of 
$(D-2)$-dimensional spacelike surfaces $S_{X^a}$ can be described by 
$\{x^a=X^a\}$, with $a,b,\dots =0,1$, where $X^a$ are arbitrary constants
and $\{x^{\alpha}\}$ are local
coordinates in $\espaitemps$. The line-element can be written as
\begin{equation}
ds^2=g_{ab} dx^adx^b+2g_{aA}dx^adx^A+g_{AB}dx^Adx^B
\label{sdg}
\end{equation}
where $g_{\mu\nu}(x^{\alpha})$ and $\det g_{AB} >0$.
There remains the freedom
\be
x^a \longrightarrow x'^a=f^a(x^b), \,\,
x^A \longrightarrow x'^A=f^A(x^B,x^c) \label{coord}
\ee
keeping the form (\ref{sdg}) and the chosen family of surfaces. 
Using coordinate conditions one can try to achieve  
$g_{aA}=0$ or other similar simplifications, which are in fact useful in many 
applications, but I prefer to keep the full generality.

Let us calculate the scalar $\kappa$ of (\ref{kappa}) for the
surfaces $S_{X^a}$. Clearly, the imbedding (\ref{imm}) for these surfaces is
given locally by
\bean
x^a=X^a=\mbox{const.}, \hspace{.5cm} x^A=\lambda^A ,
\eean
and therefore the first fundamental form (\ref{fff}) for each $S_{X^a}$
is $\gamma _{AB}=g_{AB}(X^a,\lambda^C)$.
The null normals to each $S_{X^a}$ satisfying (\ref{k})
can be chosen as
\begin{equation}
\bm{k}^{\pm}=k^{\pm}_bdx^b\vert_{S_{X^a}}, \hspace{3mm}
g^{ab}k^{\pm}_ak^{\pm}_b\vert_{S_{X^a}}=0, \hspace{3mm}
g^{ab}k^{+}_ak^{-}_b\vert_{S_{X^a}}=-1 \label{kpm}
\end{equation}
($g^{ab}$ is not necessarily the inverse of $g_{ab}$!).
Now the calculation of the $K^{\pm}_{AB}$ in (\ref{sff})
is straightforward:
\begin{eqnarray*}
K^{\pm}_{AB}=-\left.k^{\pm}_c\Gamma^c_{AB}\right\vert_{S_{X^a}}
\end{eqnarray*}
where $\Gamma^{\rho}_{\mu \nu}$ are the Christoffel symbols. From its 
definition 
$2\Gamma^a_{AB}=-g^{a\rho}(2\partial_{(A}g_{B)\rho}-\partial_{\rho} g_{AB})$
so that, by setting
\be
G\equiv +\sqrt{\det g_{AB}} \equiv e^U, \hspace{3mm} \bm{g}_{a}\equiv 
g_{aA}dx^A \label{U}
\ee
and using $k^{\pm B}=-\gamma^{BA}g_{aA}k^{\pm a}$,
the two traces (\ref{tr}) can be obtained ($f_{,\mu}= \partial_{\mu} f$)
\begin{equation}
K^{\pm}_{{}_{X^a}}=\left.k^{\pm 
a}\left(\frac{G_{,a}}{G}-\frac{1}{G}(G\gamma^{AB}g_{aA})_{,B}\right)
\right\vert_{S_{X^a}} \, . \label{pref}
\end{equation}
Hence, the mean curvature one-form reads
\be
\fbox{$\displaystyle{
H_{\mu}=\delta^a_{\mu}\left(U_{,a}-\di \vec{g}_{a} \right)}$}
\label{H}
\ee
where $\di$ is the divergence operator on vectors at each $S_{X^a}$, 
so that the scalar (\ref{kappa}) for each $S_{X^a}$ is finally
\begin{equation}
\fbox{$\displaystyle{\kappa_{{}_{X^a}}=-\left.g^{bc}H_{b}H_{c}
\right\vert_{S_{X^a}}}$}\enspace .
\label{kap}
\end{equation}
(\ref{H}-\ref{kap}) are the desired formulae, which are 
invariant under changes of type (\ref{free}) and (\ref{coord}).
Observe that one only needs to compute
the norm of $H_{a}$
{\it as if it were a one-form in the 2-dimensional metric $g^{ab}$}. 
The function $G=e^U$, which from (\ref{U}) gives the canonical $(D-2)$-volume
element of the surfaces $S_{X^a}$ (their area in $D=4$),
arises as a fundamental object. As (\ref{H}) shows, $\bm{H}$ has a pure
divergence term in general.
However, as we are going to see presently, in many 
situations $\di \vec{g}_{a}=0$, in which case $H_{a}=U_{,a}$ and only 
the normal variation of volume is relevant. 
Let us stress that (\ref{H}-\ref{kap}) are purely geometric,
independent of any matter contents,
of energy or causality conditions \cite{HE,S},
and of any field equations. They hold in general dimension $D$, 
including, in particular, the case of GR for $D=4$.

In general, $H_{a}$ will change its causal 
character at different regions. The hypersurface(s) of separation ${\cal H}$,
defined locally by the vanishing of $g^{bc}H_{b}H_{c}$, is a fundamental
place in $\espaitemps$ that I call the $S_{X^a}$-{\it horizon}. This 
contains (i) the regions with marginally trapped $S_{X^a}$, and (ii) the 
parts of each $S_{X^a}$ where one of the traces vanishes. ${\cal H}$
coincides in many cases with the classical horizons, as shown in the
examples that follow.

Many interesting applications can be derived from (\ref{H}-\ref{kap}).
Let us start with a simple illustrative example, the Kerr metric in 
Boyer-Lindquist coordinates $\{t,r,\theta,\phi\}$ (notation as in \cite{HE}).
The case of physical interest arises for $\{x^a\}=\{t,r\}$, 
$\{x^A\}=\{\theta,\phi\}$. It is immediate to obtain
$\bm{g}_{r}=0$, $\bm{g}_{t}=2amr\rho^{-2}\sin^2\theta d\phi$ and 
$$
e^{2U}=\sin^2\theta [(r^2+a^2)\rho^2 +2mra^2\sin^2\theta]
$$
so $\di \vec{g}_{a}=0$, $H_{a}=U_{,a}$, and using 
(\ref{kap}) one easily derives (for $r>0$)
$\sg\, \kappa_{{}_{t,r}}=-\sg \Delta $,
with $\Delta =r^2-2mr+a^2$. This is the standard result, which  
identifies the classical event and Cauchy horizons at $\Delta =0$
as well as the {\it closed} trapped surfaces at $\Delta <0$.

Let us consider now the general spherically
symmetric line-element in arbitrary $D$
\be
ds^2=g_{ab}(x^c)dx^adx^b+R^2(x^c)\doo \label{sph}
\ee
where $\doo$ is the round metric on
the $(D-2)$-sphere and $\det g_{ab}<0$.
Here $H_{a}= U_{,a}\propto R_{,a}/R$, and ${\cal
H}$ is the classical apparent horizon \cite{HE,S}, which in
particular becomes an event/Cauchy horizon in symmetric cases. The
former case includes $D$-dimensional Robertson-Walker cosmologies, and
the latter the Reissner-Nordstr\"om-Tangherlini black holes \cite{T},
among many others. A $D$-generalization of the standard ``mass function''
in spherical symmetry (see in GR, e.g., \cite{H}) arises
\bean
2 M(x^a)\equiv R^{D-3}\left(1-g^{bc}R_{,b}R_{,c}\right) 
\eean
so that $2M(X^a)>R^{D-3}(X^a)$ for trapped $(D-2)$-spheres $S_{X^a}$,
which in this case are obviously closed.
This agrees (up to a constant $(D-2)$-volume factor) with \cite{VW}.

For more up-to-date matters, let us apply the above results to the 
5-dimensional {\em rotating black rings/holes} recently presented in 
\cite{ERl} (containing a subset of the rotating black holes in 
\cite{MP}). Using the notation in \cite{ERl} for the metric 
appearing in their formula (13), the physically relevant case arises 
for $\{x^a\}=\{\chi,y\}$, $\{x^A\}=\{v,x,\phi\}$. By simple 
inspection one reads off
$$
\bm{g}_{y}=0, \hspace{3mm} 
\bm{g}_{\chi}=\frac{\sqrt{\nu} (y-\xi_{2})}{\sqrt{\xi_{1}}A}
\frac{F(x)}{F(y)}dv ,\hspace{4mm}
e^{2U}=-\frac{F^3(y)}{A^4(x-y)^4}
$$
and a very simple calculation gives $\di \vec{g}_{\chi}=0$
(ergo $H_{a}=U_{,a}$ once again) and, for the scalar (\ref{kap})
$$
\kappa_{{}_{\chi ,y}}=\frac{A^2(x-y)^2}{4F(x)F(y)}
\left(\frac{3F'(y)}{F(y)}+\frac{4}{x-y}\right)^2\, G(y).
$$
Recalling that $F(y)F(x)<0$ one gets
$\sg \kappa_{{}_{\chi ,y}}=-\sg G(y)$ except at $\Sigma:\, 
y+3x=4\xi_{1}$, where $\kappa_{{}_{\chi ,y}}=0$. This is due to the 
vanishing of $K^{\pm}_{{}_{\chi ,y}}$ at the
{\it 2-surfaces} $(\Sigma\cap S_{\chi ,y})\subset S_{\chi ,y}$. One can check 
that $\Sigma$ is located at $y>\xi_{1}$. Thus, there are {\em closed}
trapped 3-surfaces $S_{\chi ,y}$ for some $y\in (\xi_{1},\xi_{4})$
($G(y)$ changes sign at $\xi_{4}$), while they are 
non-trapped for $y>\xi_{4}$. The horizon formed by closed marginally trapped 
surfaces is located at $y=\xi_{4}$, which is the event horizon 
described in \cite{ERl}. In general, ${\cal H}=\Sigma\cup \{y=\xi_{4}\}$.

Another interesting application arises from the ``generalized
Weyl solutions'' constructed recently in \cite{ER}. The main aim in \cite{ER}
was to obtain the $D$-generalization of the static
and axisymmetric solutions of vacuum Einstein's equations.
However, many other solutions not of Weyl-type were implicitly,
maybe inadvertently, found.
The general metric of \cite{ER} (for the ``non-Weyl''
case characterized by having real coordinates $\{Z,\bar{Z}\}$ in the notation
of \cite{ER}) can be written in the form (\ref{sdg}) by putting $g_{aA}=0$,
$g_{AB}=$diag$\{e^{2U_{2}},\dots ,e^{2U_{D-1}}\}$,
and letting $g_{ab}$ and $\{U_A\}$ to
depend only on $\{x^a\}$. As proved with a particular coordinate choice
in \cite{ER}, the Ricci-flat
condition for (\ref{sdg}) implies then that $G=e^U$ satisfies (; is covariant
derivative for $g^{ab}$)
\be
g^{ab}G_{;ab}=0, \hspace{1cm} g^{ab}\left(G\, U_{A,a}\right)_{;b}=0
\hspace{2mm} \forall A . \label{eqs}
\ee
These two expressions are conformally invariant with respect to $g^{ab}$.
The first is simply the wave equation in the 2-metric $g^{ab}$
for $G$, easily solvable in appropriate
coordinates. The second relation (\ref{eqs}) is identical with the equation in
$D=4$, so that in what follows one could always write down
for each of the $U_{A}$ the widely known solutions found in GR.
Notice, though, that the proper choice of coordinates depends on the
particular physical situation to be tackled. For instance, a
simple possibility would be
\be
g_{ab}dx^adx^b=F^2(t,x)(-dt^2+dx^2), \hspace{3mm} G=x\, .\label{1}
\ee
However, from the previous analysis, this immediately implies that
$\kappa_{{}_{t,x}}<0$, so that {\em all} the surfaces $t,x=$const.\ {\it are
absolutely non-trapped}. \footnote{Whether or not these surfaces are closed
is an open question at this stage, depending on the specific topology of
the coordinates $\{x^ A\}$.}
In other words, the choice (\ref{1}) is adequate only
for the regions of the spacetime with absolutely
non-trapped $S_{t,x}$, and without $S_{t,x}$-horizon ${\cal H}$.
Analogous cases are given  for instance by $G=\cosh t \sinh x$.
These situations are appropriate to describe cylindrical or plane symmetric
spacetimes, providing $D$-generalizations of this type of solutions in GR.
This case is mentioned in \cite{ER}. There are, however,
other physically inequivalent situations depending on the causal
character of $U_{,a}$. These are essentially the following (keeping always
the form of $g_{ab}$ in (\ref{1}) for simplicity):

1.  $G=t$. Now all surfaces $S_{t,x}$
    are trapped and again there is no ${\cal H}$. This case describes
    cosmological solutions, as for instance
    the $D$-dimensional Kasner metric \cite{IM}, given by $U_{A}=p_{A}\log t$
    with $\sum_{A}p_{A}=1$. Analogous cases are given by
    $G=\sinh t \cosh x$.

2. $G=G(t-x)$ (or $G=G(t+x)$).
    In this case $\kappa_{{}_{t,x}}=0$ and all the surfaces $S_{t,x}$
    are marginally trapped. This kind of metrics include the plane waves subset
    of the ``pp-waves'' (see e.g. \cite{K,Gr,PW,Sc} and references therein),
    although in $D>4$ they are surprisingly
    richer as shown
    in Appendix B of \cite{ER}. Still, there is no ${\cal H}$ and $S_{t,x}$
    are generically non-closed. The case with 
    $G$=const.\ is included here.

3. $G=\sin t \sin x$ with $x$
    periodic. These cases allow for topologies $S^2$, $\r\times
    S^1$, etc in the $\{x^a\}$-part of $\espaitemps$, and are
    $D$-generalizations of the Gowdy models \cite{G}, including some
    Robertson-Walker cosmologies. In this case
    there is a non-trivial $S_{t,x}$-horizon ${\cal H}$, with two
    connected components, which splits the spacetime into 4 regions,
    two of them with trapped surfaces $S_{t,x}$ (which may be closed),
    the other two without
    them. A similar but open-universe case arises by setting
    $G=\sinh t \sinh x$.

4. Of course, one can use the general solution
    $G=f_{1}(t-x)+f_{2}(t+x)$, with arbitrary functions
    $f_{1},f_{2}$. This is specially appropriate, with adequate
    choices of $f_{1},f_{2}$, to describe the collision of
    plane waves (see e.g. \cite{Gr} and references therein for the
    GR case; $D$-generalizations were given in
    \cite{Sc,BV}). The standard procedure to build the colliding-wave
    spacetime is to replace $f_1$ and $f_2$ by
    $f_{1}(\Theta(t-x)),f_{2}(\Theta(t+x))$, where $\Theta$ is the
    Heaviside step function. Then, the two regions with $(t-x)(t+x)<0$
    are two plane waves, and the two
    zones with $(t-x)(t+x)>0$ correspond to their interaction
    region and the flat background.

 From the above one can derive yet another application. As is known,
the previous vacuum solutions can be seen as Kaluza-Klein, or
string/M-theory spacetimes which under dimensional reduction become
4-dimensional spacetimes with a number of scalar fields. The scalar
fields are given by a subset of the $\{U_{A}\}$
or appropriate linear combinations of them, see e.g \cite{Dto4,FV}. Actually,
there is a (apparently overlooked) one-to-one corespondence between the
solutions found in \cite{ER} and the solutions
generated by the technique explained in section 2 of \cite{FV}, the
$\{U_{A}\}$ of the former corresponding to the $\{p,\psi_{i}\}$ of the latter,
as can be easily proved.
One of the simplest dimensional reductions \cite{Dto4,FV}
starts with a line-element of type
\bean
ds^2_{D}=\exp{\left(-\sum_{i=4}^{D-1}\psi_{i}\right)}\, ds_{4}^2+
\sum_{i=4}^{D-1}e^{2\psi_{i}}(dx^i)^2 
\eean
where $ds_{4}^2$ is a 4-dimensional line-element and $\{x^i\}$ are
coordinates on a $(D-4)$-torus. As is clear, the physically 
observable $(D-2)$-surfaces are those reduced to 2-surfaces in
$ds_{4}^2$. For these, $G^2=\det g_{AB}$ and $\bm{g}_{a}$ become
simply ($A',B',\dots =2,3$)
\bean
G^2=\exp{\left(-2\sum_{i=4}^{D-1}\psi_{i}\right)}(\det g_{A'B'})
\prod_{i=4}^{D-1}e^{2\psi_{i}}=\det g_{A'B'}\\
\bm{g}_{a}=\exp{\left(-\sum_{i=4}^{D-1}\psi_{i}\right)}
g_{aA'}dx^{A'}\Longrightarrow \,\,
g_{a}^A=\delta_{A'}^{A}\gamma^{A'B'}g_{aB'}
\eean
so that the trapping properties of the surfaces $S_{X^a}$ {\it remain
unchanged} whether they are seen as 2-surfaces in $D=4$ or as
$(D-2)$-surfaces in full $D$. Hence, the $S_{X^a}$-horizon ${\cal H}$ is lifted
(or reduced!) from 4 to $D$ ($D$ to 4) dimensions. The considered
surfaces are closed in $D$ dimensions if and only if they are closed in $D=4$.

Let us come back to the theoretical approach. The results can be strengthened
to the case of {\it closed} trapped
surfaces. Take {\it any} closed spacelike $(D-2)$-surface $\bar{S}$ 
and adapt the coordinates such that
$\bar{S}\subset \Sigma_{f}\equiv \left\{f(x^a)=0\right\}$
where $\d f$ is timelike everywhere (apart from this $f$ is arbitrary).
This can be done in many different ways.
The imbedding (\ref{imm}) for $\bar{S}$ is given by
\be
x^a=\Phi^a(\mu^C), \hspace{.2cm}
x^A=\Phi^A(\mu^C), \hspace{2mm}
f\left(\Phi^a(\mu^C)\right)=0 , \label{im2}
\ee
where $\{\mu^C\}$ are intrinsic coordinates on $\bar{S}$. As
$\bar{S}$ is compact without boundary, $\Phi^a$ must reach their
maximum and minimum somewhere on $\bar{S}$. From
(\ref{im2}) it follows that $f_{,a}\partial \Phi^a/\partial \mu^C
=0$ so that, at any point $q\in \bar{S}$ where there is an extreme of $\Phi^1$
($\partial \Phi^1/\partial \mu^C\vert_q =0$), $\Phi^0$ must have
a critical point ($\partial \Phi^0/\partial \mu^C\vert_q =0$).
This implies, firstly, that the two future-directed null normals
of $\bar{S}$ at $q$ are given by $\bm{k}^{\pm}$ of (\ref{kpm}), and secondly,
that $\det (\partial \Phi^{C}/\partial \mu^A)\vert_{q} \neq 0$
(so that the imbedding (\ref{im2}) has rank
$(D-2)$ there.) A way to visualize this is
that $\bar{S}$ must be bi-tangent to some $S_{X^a}$
(here the assumption that
the surfaces $S_{X^a}$ do not intersect is needed, so that $\Phi^a$ 
are differentiable.) Consequently, we can choose
$\{\mu^C\}$ in a neighbourhood $V(q)\subset\bar{S}$ of $q$ such
that $\Phi^A(\mu^C)=\mu^A$ at $V(q)$.
A straightforward calculation for the traces
(\ref{tr}) of $\bar{S}$ leads now to
\begin{equation}
\left.K^{\pm}_{{}_{\bar{S}}}\right\vert_q=\left.K^{\pm}_{{}_{X^a}}\right\vert_q
-\left. k^{\pm}_{a}\gamma^{AB}_{{}_{\bar{S}}}\frac{\partial^2 \Phi^a}
{\partial \mu^A \partial \mu^B}\right\vert_q \label{pref3}
\end{equation}
where $K^{\pm}_{{}_{X^a}}$ are given in (\ref{pref}) and
$\gamma^{AB}_{{}_{\bar{S}}}$ is
the contravariant first fundamental form of $\bar{S}$.
The analysis of the second term in the righthand side of (\ref{pref3})
can be done as follows. Due to (\ref{im2})
$$
f_{,a}\frac{\partial^2 \Phi^a}
{\partial \mu^A \partial \mu^B}\vert_{q}=0\, \Longrightarrow \hspace{2mm}
f_{,a}\gamma^{AB}_{{}_{\bar{S}}}\frac{\partial^2 \Phi^a}
{\partial \mu^A \partial \mu^B}\vert_{q}=0,
$$
and given that the 
gradient of $f$ is timelike everywhere, it follows that 
$\gamma^{AB}_{{}_{\bar{S}}}\frac{\partial^2 \Phi^a}
{\partial \mu^A \partial \mu^B}\vert_q$ is spacelike with respect to 
$g^{ab}$. Hence, as the inverse of $g^{ab}$ is $-2k^+_{(a}k^-_{b)}$, both
$k^{\pm}_{a}\gamma^{AB}_{{}_{\bar{S}}}\frac{\partial^2 \Phi^a}
{\partial \mu^A \partial \mu^B}\vert_q$ must have opposite signs.

Some interesting conclusions can be derived. Assume that $H_{a}$ is spacelike
in a region, then one of the $K^{\pm}_{{}_{X^a}}$
is positive and the other is negative for all values of $X^a$
in that region. Letting $K^{+}_{{}_{X^a}}>0$ (say), then at the maximum or the
minimum of $\Phi^1$ one can always show that
\begin{eqnarray*}
\left.K^{+}_{{}_{\bar{S}}}\right\vert_q\geq
\left.K^{+}_{{}_{X^a}}\right\vert_q > 0 ,
\hspace{1cm}
\left.K^{-}_{{}_{\bar{S}}}\right\vert_q\leq
\left.K^{-}_{{}_{X^a}}\right\vert_q < 0
\end{eqnarray*}
and thus the surface $\bar{S}$ is untrapped. In fact, from the
above reasoning follows the sufficiency that neither of the
$K^{\pm}_{{}_{X^a}}$ changes sign. Thus, there are
no closed trapped surfaces at the hypersurfaces $\Sigma_f$ in any region
where the $S_{X^a}$ are marginally trapped or absolutely non-trapped.

The last result has many applications too. As a simple but
powerful one, consider once again the spherically symmetric
line-element (\ref{sph}) and assume that $R_{,\mu}$ is non-timelike everywhere.
Then, the $(D-2)$-spheres are either absolutely non-, or marginally, trapped
everywhere. Besides, this implies that {\it there cannot be any closed trapped
surface in $\espaitemps$ at all}.  For, due to the structure of the
manifold, any closed $\bar{S}$ must be osculating to some
$(D-2)$-sphere somewhere, and at this point their corresponding
$\kappa$'s of (\ref{kap}) coincide, proving that $\bar{S}$ cannot
be trapped there.
In particular, {\it globally} static cases have $R_{,\mu}$
non-timelike everywhere, 
hence {\it they cannot contain closed trapped surfaces}. These include
flat spacetime, or Einstein's and anti-de Sitter's universes, for example.
Of course, this is a known result, but the proof was rather indirect:
if there were a closed trapped surface, the spacetime would be
geodesically incomplete \cite{P2,HE,S}, which is not.
It must be noted that any of the mentioned spacetimes, including flat 
one, certainly contains trapped surfaces (non-compact!, see \cite{S} for
examples), so that the previous result is not obvious in principle.

All in all, the significance and applicability of (\ref{H}-\ref{kap}), 
which are wholly general, extremely simple, and easily computable,
seems worth exploring in detail in the many different 
gravitational theories. Their potential applications seem to be very 
many.

\vspace{5mm}
\noindent
{\bf Acknowledgement}

\vspace{5mm}
\noindent
I thank Roberto Emparan, Alex Feinstein, Marc Mars, and Ra\"ul Vera 
for their comments and suggestions.

\end{document}